\documentclass[aps,superscriptaddress,tightenlines,nofootinbib,floatfix,longbibliography,notitlepage,twocolumn]{revtex4-1}
\usepackage[left=18mm,right=19mm,top=23mm,bottom=16mm]{geometry}
\usepackage{placeins}
\usepackage{bbm}
\usepackage{graphicx}
\usepackage[T1]{fontenc}
\usepackage[utf8]{inputenc}
\usepackage[UKenglish]{babel}
\usepackage{siunitx}
\usepackage[dvipsnames]{xcolor}
\usepackage{listings}
\usepackage{amsmath}
\usepackage[super]{nth}
\usepackage{mathtools}
\usepackage{amssymb}
\usepackage{dsfont}
\usepackage{upgreek}
\usepackage{tikz-feynman}
\usepackage{tikz}
    \usetikzlibrary{arrows,matrix,arrows.meta,shapes}
    \tikzstyle{arrow} = [very thick,->,>=stealth]
\usepackage{braket}
\usepackage{xfrac}
\usepackage{setspace}
\usepackage[linesnumbered,lined,boxed,commentsnumbered]{algorithm2e}
\usepackage{nameref,zref-xr}
\usepackage{hyperref}
\usepackage{cleveref}
\usepackage{comment}
\usepackage{nicefrac}
\usepackage[caption=false]{subfig}
\usepackage{amsfonts}
\zxrsetup{toltxlabel}
\zexternaldocument*{supplementary}

\graphicspath{{figure/},{data/}}

\providecommand{\repositoryInformationSetup}{} %
\repositoryInformationSetup

\newcommand{\repoURL}{https://github.com/evanberkowitz/hubbard-practical}

\newcommand{\Figref}[1]{Figure~\ref{fig:#1}\xspace}

\def\Ref#1{Ref.~\cite{#1}} %

\newcommand{\goesto}{\ensuremath{\rightarrow}}

\newcommand{\Reals}{\mathbb{R}\xspace}
\newcommand{\Complexes}{\mathbb{C}\xspace}

\newcommand{\oneover}[1]{\ensuremath{\frac{1}{#1}}}                             %
\newcommand{\inverse}{\ensuremath{^{-1}}}                                       %
\newcommand{\half}{\ensuremath{\frac{1}{2}} }                                   %
\newcommand{\abs}[1]{\ensuremath{\left| #1 \right|}\xspace}

\newcommand{\average}[1]{\ensuremath{\left\langle #1 \right\rangle}\xspace}

\DeclareMathOperator{\tr}{Tr}

\newcommand{\crit}{{\ensuremath{\textrm{c}}}}
\newcommand{\Int}{\ensuremath{\mathfrak{I}}}

\newcommand{\DD}[1]{\ensuremath{\hspace*{-3pt}\mathcal{D}#1\;}}
\newcommand{\dd}[1]{\ensuremath{\,\text{d} #1 }}
\newcommand{\NN}{\ensuremath{\mathcal{N}\mathcal{N}}}
\newcommand{\SHIFT}{\ensuremath{\text{SHIFT}}}

\newcommand{\Obs}{\ensuremath{\mathcal{O}}\xspace}
\newcommand{\Z}{\ensuremath{\mathcal{Z}}\xspace}
\newcommand{\Ncfg}{\ensuremath{N_{\textrm{conf}}}\xspace}
\newcommand{\M}{\ensuremath{\mathcal{M}}\xspace}

\let\builtinLaTeX\LaTeX
\def\LaTeX{\builtinLaTeX\xspace}

\definecolor{fzjblue}{RGB}{2,61,107} %
\colorlet{color1}{fzjblue}

\definecolor{fzjlightblue}{RGB}{173,189,227} %
\colorlet{color2}{fzjlightblue}

\definecolor{fzjgray}{RGB}{235,235,235} %
\colorlet{fzjgrey}{fzjgray}
\colorlet{color3}{fzjgray}

\definecolor{fzjred}{RGB}{235, 95, 115}  %
\colorlet{color4}{fzjred}

\definecolor{fzjgreen}{RGB}{185, 210, 95}  %
\colorlet{color5}{fzjgreen}

\definecolor{fzjyellow}{RGB}{250, 235, 90}  %
\colorlet{color6}{fzjyellow}

\definecolor{fzjviolet}{RGB}{175, 130, 185}  %
\colorlet{color7}{fzjviolet}

\definecolor{fzjorange}{RGB}{250, 180, 90}  %
\colorlet{color8}{fzjorange}

\definecolor{fzjblack}{RGB}{0,0,0}
\definecolor{fzjwhite}{RGB}{255,255,255}

\newcommand{\bonn}{
    Helmholtz-Institut f\"{u}r Strahlen- und Kernphysik,
    Rheinische Friedrich-Wilhelms-Universit\"{a}t Bonn, 53115 Bonn, Germany
}

\newcommand{\ikp}{
    Institut f\"{u}r Kernphysik,
    Forschungszentrum J\"{u}lich, 54245 J\"{u}lich, Germany
}

\newcommand{\ias}{
    Institute for Advanced Simulation,
    Forschungszentrum J\"{u}lich, 54245 J\"{u}lich, Germany
}

\newcommand{\jsc}{
    JARA \& J\"{u}lich Supercomputing Center,
    Forschungszentrum J\"{u}lich, 54245 J\"{u}lich, Germany
}

\newcommand{\liverpool}{
    Department of Mathematical Sciences,
    University of Liverpool, Liverpool, L69 7ZL, United Kingdom
}

\newcommand{\casa}{
    Center for Advanced Simulation and Analytics (CASA),
    Forschungszentrum Jülich, 52425 J\"{u}lich, Germany
}

\begin{document}

\title{Mitigating the Hubbard Sign Problem with Complex-Valued Neural Networks}

\author{Marcel Rodekamp}        \affiliation{\ias}\affiliation{\jsc}\affiliation{\casa}\affiliation{\bonn}
\author{Evan Berkowitz}         \affiliation{\ias}\affiliation{\jsc}\affiliation{\casa}
\author{Christoph Gäntgen}      \affiliation{\ias}\affiliation{\casa}\affiliation{\bonn}
\author{Stefan Krieg}           \affiliation{\ias} \affiliation{\jsc}\affiliation{\casa} \affiliation{\bonn}
\author{Thomas Luu}             \affiliation{\ias} \affiliation{\bonn} \affiliation{\ikp}
\author{Johann Ostmeyer}        \affiliation{\liverpool}

\date{\today}

\begin{abstract}
Monte Carlo simulations away from half-filling suffer from a sign problem that can be reduced by deforming the contour of integration.
Such a transformation, which induces a Jacobian determinant in the Boltzmann weight, can be implemented using neural networks.
This additional determinant cost for a generic neural network scales cubically with the volume,
preventing large-scale simulations.
We implement a new architecture, based on complex-valued affine coupling layers, which reduces this to linear scaling.
We demonstrate the efficacy of this method by successfully applying it to systems of different size, 
the largest of which is intractable by other Monte Carlo methods due to its severe sign problem.
\end{abstract}

 \maketitle

\section{Introduction}\label{sec:Introduction}
The computational sign problem encumbers successful importance sampling from complex-valued distributions with Markov Chain Monte Carlo algorithms such as Hybrid Monte Carlo (HMC).
Sampling from the configuration space of a wide variety of interesting physical systems suffers such a difficulty, ranging from lattice QCD at finite baryon chemical potential and doped condensed matter systems in equilibrium to the real-time evolution of quantum systems.

By deforming the real manifold of integration for a path integral of interest into complex variables, one may reduce the sign problem substantially~\cite{Kashiwa:2018vxr,alexandru2020complex,Detmold:2020ncp,Detmold:2021ulb}.
In the last few years, new formal developments have inspired investigation into leveraging Lefschetz thimbles~\cite{Lefschetz1921,PhysRevD.93.014504,Cristoforetti:2014gsa,Cristoforetti:2013wha,Mukherjee:2013aga,Kanazawa:2014qma,Tanizaki:2016lta}---high-dimensional analogues of contours of steepest descent which can be located by holomorphic flow.
In~\cite{PhysRevD.93.014504}, for example, fluctuations about the saddle point of each thimble were sampled to simulate the 0+1 dimensional Thirring model, something much akin to the method of steepest descent.
In practice the determination of the precise location of each thimble's saddle point, or critical point, as well as the relevant sampling `direction' about these points, is numerically costly and prohibitive.  
An alternative method is to train neural networks to learn the map from some starting manifold to any beneficial manifold, including one that approximates the thimbles that contribute to the integral~\cite{PhysRevD.96.094505,Mori:2017nwj,leveragingML}.

In our previous work~\cite{leveragingML} we were limited by the computational cost of incorporating the Jacobian determinant of this map into our importance sampling.
In this paper we leverage complex-valued neural networks built of affine coupling layers to reduce the scaling of the Jacobian determinant cost.
We focus on the Hubbard model on a honeycomb lattice away from half-filling and compare methods by computing single-particle correlation functions.

This paper is organized in the following way.
In Section \ref{sec:Formalism}, a brief recap of the Hubbard model and basic notation is given.
After that, some prior methods to alleviate or remove the sign problem and usage within HMC are discussed.
In Section \ref{sec:Method}, we describe the new neural network architecture.
In Section \ref{sec:Results}, we show a numerical test of the network on three systems where we can exactly diagonalize the Hamiltonian, and one larger system beyond our ability to exactly diagonalize.

\section{Formalism}\label{sec:Formalism}
\begin{figure}
    \subfloat[2 Sites]{%
        \hspace*{1em}
        \raisebox{1.2em}{
            \begin{tikzpicture}
                \draw[draw=fzjblue!109!fzjblack, line width=0.5, line cap=round] (-0.5,0) -- (0.5,0);
                \draw[ball color=fzjviolet] (-0.5,0) circle (3pt);
                \draw[ball color=fzjviolet] (0.5,0) circle (3pt);
            \end{tikzpicture}
        } %
        \hspace*{1em}
    } %
    \subfloat[4 Sites]{%
        \raisebox{0.2em}{
            \resizebox{0.3\linewidth}{!}{
                \begin{tikzpicture}
                    \draw[draw=fzjblue!109!fzjblack, line width=0.5, line cap=round] (0,0) -- (1,0);
                    \draw[draw=fzjblue!109!fzjblack, line width=0.5, line cap=round, dashed] (0,0) -- (2.5,0.866025403784439);
                    \draw[draw=fzjblue!109!fzjblack, line width=0.5, line cap=round] (1,0) -- (1.5,0.866025403784439);
                    \draw[draw=fzjblue!109!fzjblack, line width=0.5, line cap=round] (1.5,0.866025403784439) -- (2.5,0.866025403784439);
                    \draw[ball color=fzjviolet] (0,0) circle (3pt);
                    \draw[ball color=fzjviolet] (1,0) circle (3pt);
                    \draw[ball color=fzjviolet] (1.5,0.866025403784439) circle (3pt);
                    \draw[ball color=fzjviolet] (2.5,0.866025403784439) circle (3pt);
                \end{tikzpicture}
            } %
        } %
    } %
    \subfloat[8 Sites]{
        \resizebox{0.3\linewidth}{!}{
            \begin{tikzpicture}
                \draw[draw=fzjblue!109!fzjblack, line width=0.5, line cap=round] (0,0) -- (1,0);
                \draw[draw=fzjblue!109!fzjblack, line width=0.5, line cap=round, dashed] (0,0) -- (2.5, 0.866025403784439);
                \draw[draw=fzjblue!109!fzjblack, line width=0.5, line cap=round, dashed] (0,0) -- (2.5, -0.866025403784439);
                \draw[draw=fzjblue!109!fzjblack, line width=0.5, line cap=round] (1,0) -- (1.5, 0.866025403784439);
                \draw[draw=fzjblue!109!fzjblack, line width=0.5, line cap=round] (1,0) -- (1.5, -0.866025403784439);
                \draw[draw=fzjblue!109!fzjblack, line width=0.5, line cap=round] (1.5, 0.866025403784439) -- (2.5, 0.866025403784439);
                \draw[draw=fzjblue!109!fzjblack, line width=0.5, line cap=round, dashed] (1.5, 0.866025403784439) -- (4,0);
                \draw[draw=fzjblue!109!fzjblack, line width=0.5, line cap=round] (2.5, 0.866025403784439) -- (3,0);
                \draw[draw=fzjblue!109!fzjblack, line width=0.5, line cap=round] (1.5,-0.866025403784439) -- (2.5,-0.866025403784439);
                \draw[draw=fzjblue!109!fzjblack, line width=0.5, line cap=round, dashed] (1.5,-0.866025403784439) -- (4,0);
                \draw[draw=fzjblue!109!fzjblack, line width=0.5, line cap=round] (2.5,-0.866025403784439) -- (3,0);
                \draw[draw=fzjblue!109!fzjblack, line width=0.5, line cap=round] (3,0) -- (4,0);

                \draw[ball color=fzjviolet] (0, 0) circle (3pt);
                \draw[ball color=fzjviolet] (1, 0) circle (3pt);
                \draw[ball color=fzjviolet] (1.5, 0.866025403784439) circle (3pt);
                \draw[ball color=fzjviolet] (2.5, 0.866025403784439) circle (3pt);
                \draw[ball color=fzjviolet] (1.5, -0.866025403784439) circle (3pt);
                \draw[ball color=fzjviolet] (2.5, -0.866025403784439) circle (3pt);
                \draw[ball color=fzjviolet] (3,0) circle (3pt);
                \draw[ball color=fzjviolet] (4,0) circle (3pt);
            \end{tikzpicture}
        } %
    } %
    \\
    \subfloat[18 Sites (boundary suppressed)]{\qquad
        \resizebox{0.4\linewidth}{!}{
        \begin{tikzpicture}
    \def\i{0};
    \def\j{1};
    \def\spacing{0.5}
    \def\xshift{ \spacing*\i*(3/2) }
    \ifthenelse{\isodd{\i}}{
        \def\yshift{ sqrt(3)*(\j+0.5)*\spacing }
    } {
        \def\yshift{ sqrt(3)*\j*\spacing }
    }
    
    \coordinate (A) at ({\xshift+\spacing}         ,{\yshift}                  );
    \coordinate (B) at ({\xshift+\spacing*cos(60)} ,{\yshift+\spacing*sin(60)} );
    \coordinate (C) at ({\xshift+\spacing*cos(120)},{\yshift+\spacing*sin(120)});
    \coordinate (D) at ({\xshift-\spacing}         ,{\yshift}                  );
    \coordinate (E) at ({\xshift+\spacing*cos(240)},{\yshift+\spacing*sin(240)});
    \coordinate (F) at ({\xshift+\spacing*cos(300)},{\yshift+\spacing*sin(300)});
    
    \draw[draw=fzjblue!109!fzjblack, line width=0.5, line cap=round] (A) -- (B);
    \draw[draw=fzjblue!109!fzjblack, line width=0.5, line cap=round] (B) -- (C);
    \draw[draw=fzjblue!109!fzjblack, line width=0.5, line cap=round] (C) -- (D);
    \draw[draw=fzjblue!109!fzjblack, line width=0.5, line cap=round] (D) -- (E);
    \draw[draw=fzjblue!109!fzjblack, line width=0.5, line cap=round] (E) -- (F);
    \draw[draw=fzjblue!109!fzjblack, line width=0.5, line cap=round] (F) -- (A);
    
    \coordinate (dangleLeft1) at ({\spacing*-1*(3/2)-\spacing},{sqrt(3)*(0.5)*\spacing});
    \coordinate (dangleLeft2) at ({\spacing*-1*(3/2)-2*\spacing},{sqrt(3)*(0.5)*\spacing});
    \draw[draw=fzjblue!109!fzjblack, line width=0.5, line cap=round] (dangleLeft1) -- (dangleLeft2);
    \coordinate (dangleLeft3) at ({\spacing*1*(3/2)+\spacing},{sqrt(3)*(0.5)*\spacing});
    \coordinate (dangleLeft4) at ({\spacing*1*(3/2)+2*\spacing},{sqrt(3)*(0.5)*\spacing});
    \draw[draw=fzjblue!109!fzjblack, line width=0.5, line cap=round] (dangleLeft3) -- (dangleLeft4);
        \foreach \j in {0}{%
            \foreach \i in {-1,0,1}{%
            \def\spacing{0.5}
            \def\xshift{ \spacing*\i*(3/2) }
            \ifthenelse{\isodd{\i}}{
                \def\yshift{ sqrt(3)*(\j+0.5)*\spacing }
            } {
                \def\yshift{ sqrt(3)*\j*\spacing }
            }
            
            \coordinate (A) at ({\xshift+\spacing}         ,{\yshift}                  );
            \coordinate (B) at ({\xshift+\spacing*cos(60)} ,{\yshift+\spacing*sin(60)} );
            \coordinate (C) at ({\xshift+\spacing*cos(120)},{\yshift+\spacing*sin(120)});
            \coordinate (D) at ({\xshift-\spacing}         ,{\yshift}                  );
            \coordinate (E) at ({\xshift+\spacing*cos(240)},{\yshift+\spacing*sin(240)});
            \coordinate (F) at ({\xshift+\spacing*cos(300)},{\yshift+\spacing*sin(300)});

            \draw[draw=fzjblue!109!fzjblack, line width=0.5, line cap=round] (A) -- (B);
            \draw[draw=fzjblue!109!fzjblack, line width=0.5, line cap=round] (B) -- (C);
            \draw[draw=fzjblue!109!fzjblack, line width=0.5, line cap=round] (C) -- (D);
            \draw[draw=fzjblue!109!fzjblack, line width=0.5, line cap=round] (D) -- (E);
            \draw[draw=fzjblue!109!fzjblack, line width=0.5, line cap=round] (E) -- (F);
            \draw[draw=fzjblue!109!fzjblack, line width=0.5, line cap=round] (F) -- (A);

            \draw[ball color=fzjviolet] (A) circle (3pt);
            \draw[ball color=fzjviolet] (B) circle (3pt);
            \draw[ball color=fzjviolet] (C) circle (3pt);
            \draw[ball color=fzjviolet] (D) circle (3pt);
            \draw[ball color=fzjviolet] (E) circle (3pt);
            \draw[ball color=fzjviolet] (F) circle (3pt);
        }%
    }%

    \def\i{0};
    \def\j{1};
    \def\spacing{0.5}
    \def\xshift{ \spacing*\i*(3/2) }
    \ifthenelse{\isodd{\i}}{
        \def\yshift{ sqrt(3)*(\j+0.5)*\spacing }
    } {
        \def\yshift{ sqrt(3)*\j*\spacing }
    }
    
    \coordinate (A) at ({\xshift+\spacing}         ,{\yshift}                  );
    \coordinate (B) at ({\xshift+\spacing*cos(60)} ,{\yshift+\spacing*sin(60)} );
    \coordinate (C) at ({\xshift+\spacing*cos(120)},{\yshift+\spacing*sin(120)});
    \coordinate (D) at ({\xshift-\spacing}         ,{\yshift}                  );
    \coordinate (E) at ({\xshift+\spacing*cos(240)},{\yshift+\spacing*sin(240)});
    \coordinate (F) at ({\xshift+\spacing*cos(300)},{\yshift+\spacing*sin(300)});

    \draw[ball color=fzjviolet] (A) circle (3pt);
    \draw[ball color=fzjviolet] (B) circle (3pt);
    \draw[ball color=fzjviolet] (C) circle (3pt);
    \draw[ball color=fzjviolet] (D) circle (3pt);
    \draw[ball color=fzjviolet] (E) circle (3pt);
    \draw[ball color=fzjviolet] (F) circle (3pt);
    \draw[ball color=fzjviolet] (dangleLeft2) circle (3pt);
    \draw[ball color=fzjviolet] (dangleLeft4) circle (3pt);

    \end{tikzpicture}
    } %
\qquad
    } %
    \caption{
        Graphical representation of the arrangement of ions considered in the numerical investigation.
        Each node corresponds to an ion while each edge indicates an allowed particle/hole hopping.
        The dashed lines represent the periodic boundary.
    }\label{fig:models}
\end{figure}
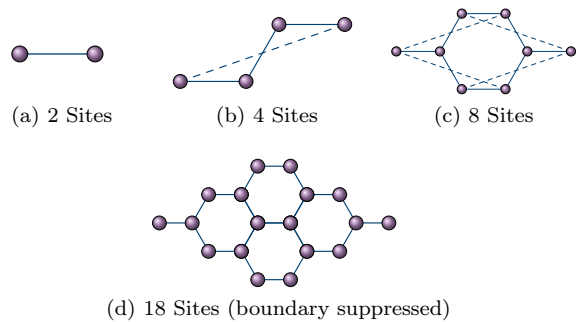

The Hubbard model~\cite{Hubbard1963} describes a fixed spatial lattice $X$ on which particles can move and interact.
In the particle-hole basis it is described by Hamiltonian
\begin{align}
    \nonumber
    \mathcal{H}\left[K, V, \mu\right]
    =
        &   - \sum_{x,y\in X} \left( p_x^\dagger K^{xy} p_y - h_x^\dagger K^{xy} h_y \right)
    \\  &   + \frac{1}{2} \sum_{x,y\in X} \rho_x V^{xy} \rho_y
            + \mu \sum_{x\in X} \rho_x,
    \label{eq:hubbard-hamiltonian}
\end{align}
where the amplitudes in $K$ encode the hopping of fermionic particles $p$ and holes $h$, the potential $V$ describes the interactions between charges
\begin{equation}
    \rho_x = p^{\dagger}_x p_x - h^{\dagger}_x h_x
    \label{eq:net-charge}
\end{equation}
and the chemical potential $\mu$ incentivizes charge.
By adjusting $K$ and $V$ this model can describe a wide variety of physical systems.
We restrict our attention to the case where $K$ encodes a honeycomb structure with nearest-neighbor hopping and the interaction $V$ is local,
\begin{align}
    K &= \kappa \delta_{\langle xy \rangle}
    &
    V &= U \delta_{xy};
\end{align}
the bipartiteness of the honeycomb permits a signed sublattice transformation that flips the sign of the hopping of holes.
As we are focusing on algorithmic issues we focus on only the four systems displayed in Figure~\ref{fig:models}.
These---the 2, 4, 8, and 18 site models---are examples of the honeycomb lattice with periodic boundary conditions.

Our aim is to compute observables \Obs according to the thermal trace
\begin{equation}
    \average{\Obs} = \oneover{\Z} \tr{\left[\Obs e^{-\beta H}\right]}.
    \label{eq:average}
\end{equation}
where the partition function \Z is the trace without the observable and $\beta$ is the inverse temperature, the euclidean time extent.
Trotterizing into $N_t$ timeslices, inserting Grassmannian resolutions of the identity, and linearizing the interaction via the Hubbard-Stratonovich transformation~\cite{Hubbard1959} leads to the action
\begin{align}
    \nonumber
    S\left[\Phi \,\vert \, K, V, \mu \right]
       =&
            -   \log\det{ M\left[\Phi\,\vert\, K,\mu\right] \cdot M\left[-\Phi\,\vert\, -K,-\mu\right] }
    \\  &   +   \frac{1}{2} \sum_{t}\sum_{x,y\in X} \Phi_{tx} (\delta V\inverse)^{xy} \Phi_{ty},
    \label{eq:hubbard-action}
\end{align}
where $\Phi \in \Reals^{\abs{\Lambda}}$ is an auxiliary field on the spacetime lattice $\Lambda = [0, N_t-1]\otimes X$ and $\delta=\beta/N_t$.
We use the exponential discretization~\cite{Wynen:2018ryx} for the fermion matrices
\begin{align}
    M\left[\Phi\,\vert\, K,\mu\right]_{x't';xt} 
    =& 
        \delta_{x'x}\delta_{t't}
    \\\nonumber
    &-  \left( e^{\delta(K + \mu)} \right)_{x'x} e^{+ i \Phi_{xt}} \mathcal{B}_{t'}\delta_{t'(t+1)}
\end{align}
where $\mathcal{B}$ encodes the antiperiodic boundary conditions in time.
On a bipartite lattice we may replace the $-K$ in the holes' fermion matrix with $+K$; then when $\mu=0$ the determinant may be made manifestly positive-semidefinite.
When $\mu$ is finite $S$ is complex; a great deal of recent effort has been made in the computational physics community to understand this case~\cite{Mukherjee:2014hsa,Fukuma:2019wbv,Ulybyshev:2019fte,Schneider:2021}.

The transformation of the thermal average \eqref{eq:average} leads to the path integral
\begin{equation}
    \average{\Obs}
    =
    \oneover{\Z} \int  \DD{\Phi} e^{-\beta S\left[\Phi\right]} \Obs\left[\Phi\right]
    \equiv
    \int \DD{\Phi} p_S\left[\Phi\right] \Obs\left[\Phi\right]\label{eq:true-expectation-value}
\end{equation}
where the partition function $\Z$ is the integral without the observable \Obs.
When the action is real importance-sampling methods draw \Ncfg configurations according to the Boltzmann distribution
\begin{equation}
    p_S\left[\Phi\right] = \oneover{\Z} e^{- S\left[\Phi\right]}
    \label{eq:probability-density}
\end{equation}
and estimate observables \eqref{eq:true-expectation-value} by an unweighted average.
Any practical calculation samples only finitely many configurations $\Ncfg$ and the resulting statistical uncertainties scale like $\Ncfg^{-1/2}$ as long as the configurations are independent.

At finite $\mu$ a complex-valued action yields an oscillating integrand and $p_S$ \eqref{eq:probability-density} can no longer be interpreted as a standard probability density, rendering a straightforward application of importance sampling impossible.

To recover an importance-sampling algorithm we can separate the real and imaginary parts of the action $S=\Re{S}+i\Im{S}$
and rewrite the partition function
\begin{equation}
    \Z = \int \DD{\Phi} e^{-S} = \int \DD{\Phi} e^{-\Re{S}} e^{-i\Im{S}} \propto \average{e^{-i\Im{S}}}_{\Re{S}} \equiv \Sigma
    \label{eq:statistical-power}
\end{equation}
where the expectation value is with respect to the real part of the action and we call $\Sigma$ the statistical power.
So, by sampling according to $p_{\Re{S}}$ we can estimate
\begin{align}
        \average{\Obs}
        &   = \frac{\average{e^{-i\Im{S}}\Obs}_{\Re{S}}}{\average{e^{-i \Im{S}}}_{\Re{S}}}
            = \oneover{\Sigma} \average{ e^{-i \Im{S}} \Obs}_{\Re{S}}.
        \label{eq:reweighting}
\end{align}
When the statistical power $\Sigma$ \eqref{eq:statistical-power} cannot be reliably distinguished from zero the sign problem is too strong and the whole procedure fails~\cite{berger2021complex,leveragingML,PhysRevD.93.014504,mori2018lefschetz}.
\Ref{berger2021complex} showed that the effective number of configurations
\begin{equation}
    \Ncfg^{\textrm{eff}} = \abs{\Sigma}^2 \cdot \Ncfg \label{eq:effective-Nconf}
\end{equation}
controls the scaling of statistical errors $\sim \left(\Ncfg^{\textrm{eff}}\right)^{-1/2}$.

It is widely expected that the statistical power shrinks exponentially with spacetime volume $\beta\abs{X}$.
Because the power is the ratio of the full and phase-quenched partition functions it should be exponential in a difference of free energies, which is extensive in the spacetime volume~\cite{Splittorff:2006fu}.
For small nonbipartite examples we have previously confirmed the exponential dependence on $\beta$~\cite{leveragingML}.

A promising alternative to simple reweighting is to complexify the domain of integration and transform $\phi\in\M_{\Reals} = \Reals^{\abs{\Lambda}}$ to a manifold $\Phi\in\M \subset \Complexes^{\abs{\Lambda}}$.
As long as \M is in the same homology class, the analogue of the Cauchy integral theorem ensures that the partition function is unchanged~\cite{alexandru2020complex},
\begin{align}
    \Z = \int_\M \DD{\Phi} e^{-S[\Phi]}.
\end{align}
Parametrizing the manifold $\M$ by the real fields induces a Jacobian determinant, yielding~\cite{PhysRevD.93.014504}
\begin{align}
    \Z
    &=
    \int_{\M_\Reals} \DD{\phi} e^{-S\left[\Phi\left(\phi\right)\right] + \log{\det{J\left[\Phi\left(\phi\right)\right]}}}
    \label{eq:pullback}
\end{align}
and observables are computed on the manifold $\Obs\left[\Phi\left(\phi\right)\right]$.

A judicious choice of the manifold $\M$ can diminish or completely remove the sign problem~\cite{PhysRevD.86.074506,alexandru2020complex}.
Even when sampling according to $p_{\Re S^\textrm{eff}}$ with an imperfect manifold with a complex effective action
\begin{align}
    S^{\textrm{eff}}\left[\phi\right]
    &= S\left[\Phi(\phi)\right]
    - \log\det J\left[\Phi(\phi)\right]
    &
    J_{ij} = \frac{\partial \Phi_i}{\partial \phi_j}
    \label{eq:effective-action}
\end{align}
if the statistical power  $\Sigma$~\eqref{eq:statistical-power} is sufficiently improved we can reweight \eqref{eq:reweighting} with the imaginary part $\Im S^{\textrm{eff}}$.

There are many strategies for picking target manifolds~\cite{Tanizaki:2017yow}.
One choice is to try to approximate the Lefschetz thimbles -- high-dimensional manifolds analogous to contours of steepest descent, which have constant imaginary action and therefore have a much-reduced sign problem~\cite{PhysRevD.86.074506}.
Each thimble contains a critical point $\Phi_\crit$ that satisfies
\begin{equation}
    \left.\frac{\partial S\left[\Phi\right]}{\partial \Phi}\right|_{\Phi = \Phi_\crit} = 0
    \label{eq:critical-point}
\end{equation}
and is therefore a fixed point of the holomorphic flow
\begin{equation}
    \frac{d\Phi(\tau)}{d\tau} = \left(\frac{\partial S\left[ \Phi(\tau) \right]}{\partial \Phi(\tau)}\right)^*
    \label{eq:holomorphic-flow}
\end{equation}
as a function of the fictitious flow time $\tau$ and initial condition $\Phi(0) = \phi$.
We can trace trajectories under the flow using the integrator
\begin{equation}
    \Int_\tau^\pm\left[\phi\right] \equiv \int\limits_{0}^{\pm\tau} \left(\frac{\partial S\left[ \Phi(\tau_f) \right]}{\partial\Phi(\tau_f)}\right)^* \dd{\tau_f}.
    \label{eq:formal-integrator}
\end{equation}
A thimble is the set of complexified configurations that flow to a critical point under downward flow $\Int_\infty^-$.

There may be many thimbles in $\Complexes^{\abs{\Lambda}}$ and only some might contribute.
The upward flow $\Int_\infty^+$ discovers these thimbles automatically.
After enough flow time $\tau$ the integrator $\Int_\tau^+$ drives any $\Phi(0)$ to either a place on a thimble or to neverland -- any place where thimbles of different imaginary action meet and therefore must have zero weight.
When $\Phi$ starts on a valid integration manifold its image under $\Int^+_\infty$ is on a thimble that contributes to the integral or is in neverland.
For an approachable discussion and proof, see the recent review \Ref{alexandru2020complex}.

Therefore, we can try to evaluate the path integral \eqref{eq:pullback} on the manifold given by $\Phi\left(\phi\right) = \Int_\infty^+[\phi]$ for each $\phi$ on any valid starting manifold $\M_0$, such as $\M_\Reals$.
Though this seems to make sign problem free simulations possible, two issues remain.
While integrating the flow \eqref{eq:formal-integrator} is cheap, performing molecular dynamics integration on the thimbles at first glance involves the costly computation of the Hessian $\partial_{\Phi_i}\partial_{\Phi_j} S\left[\Phi\right]$ due to the appearance of the Jacobian determinant of the flow in the effective action~\eqref{eq:effective-action}, though some ideas for quickly estimating the Jacobian have been proposed~\cite{Alexandru:2016lsn} and recent work~\cite{Fujisawa:2021hxh} shows how to accelerate this for sparse, local (bosonic) actions.
The Jacobian determinant has to be evaluated at any accept-reject step with computational cost scaling like~$\abs{\Lambda}^3$.

Second, because thimbles only touch at places of zero weight, algorithms like HMC~\cite{Duane1987} which use a smooth update of the fields $\Phi$ would be encumbered by an ergodicity problem.
The severity of this issue is ameliorated in two ways.
As any practical integrator $\Int^+_\tau$ necessarily approximates the flow, the resulting integration manifold is only approximately the union of contributing thimbles.
Additionally, we do not need to flow for very much time.
Both of these mean that the important configurations are smoothly connected, though the imaginary part of the action is not perfectly piecewise constant.
In practice, picking a $\tau$ is a tradeoff between reducing the computational cost of the flow and an improvement of the statistical power.

The cost of the flow and the associated Jacobian determinant is such that it is beneficial to train a neural network to learn the map $\Int^+_\tau: \M_0 \goesto \tilde{\M}$.
In the next section we explain our network's architecture.

Of course, understanding neural networks as general function approximators yields an interpretation of any (numerical) integrator as a network, though it is parameter-free and needs no training---its layers, given by some discretization of the flow equations~\eqref{eq:holomorphic-flow}, are exactly known.
Just as we can produce training configurations closer to the thimbles with a more precise integrator, by adding additional layers we may train the network to reproduce the integrated flow more accurately.
So, one expects a trade-off between the nearness to the thimbles (thinking of the number of layers as a proxy) and the effort required to train.
The algorithm we describe is exact, even in the case where the network does not offer an acceleration, since the network produces a manifold with the correct homology class regardless of its fidelity to the thimbles.

Because we can integrate on any manifold in the same homology class as $\Reals^{\abs{\Lambda}}$, it may be beneficial to find simple manifolds that can improve the statistical power without the computational cost of flowing~\cite{Alexandru:2018ddf,Warrington:2019kzf}.
One such manifold is the tangent (hyper-)plane $\Phi\in\M_T$~\cite{PhysRevD.93.014504,leveragingML,alexandru2020complex}, a hyperplane parallel to the real manifold offset by a constant imaginary piece so that it intersects the critical-point image of the zero configuration $i\Phi_\crit^0 = \Int^+_\infty(0)$
\begin{align}
    \Phi\left(\phi\right) = \phi + i \Phi_\crit^0
    \label{eq:tangent-plane}
\end{align}
for all $\phi \in \M_{\Reals}$.
For many smaller systems this transformation already reduces the sign problem enough that reweighting can be applied.
However, in our larger examples the tangent plane gives no appreciable statistical power.
Nevertheless, we can reduce the cost and potentially increase the potency of flowing if we start from the tangent plane~\cite{leveragingML}.

One obvious approach to constructing an HMC-like algorithm is to attempt molecular dynamics on the target manifold $\tilde{\M}$ given by $\tilde{\Phi}$; in our case, an approximation of the thimbles.
However, remaining on the manifold is not so simple~\cite{Fujii:2013sra,Fujii:2015bua,Fukuma:2019uot,Fujisawa:2021hxh}.

In contrast, performing HMC on the tangent plane \emph{is} simple -- when integrating molecular dynamics trajectories simply neglect the imaginary part of the force.
Because the real plane suffers from a severe sign problem in the examples we study, we use this tanget-plane HMC as a benchmark.
In the remainder of this paper we refer to it simply as ``HMC'' unless clarification is needed.

For further improvement we do molecular dynamics on the tangent plane $\M_T$ and perform the Metropolis-Hastings accept/reject step on the target manifold $\tilde{\M}$ according to the effective action \eqref{eq:effective-action}.
We track the configuration on both the integration manifold $\M_0$ and its image on the target manifold $\tilde{\M}$ to avoid paying the computational cost of applying or inverting the transformation $\tilde{\Phi}$ more than needed.
Assuming the numerical implementation of the map $\tilde{\Phi}$ is invertible, proof that this algorithm has detailed balance is provided in \Ref{leveragingML}.
One can use a reversible integrator or an invertible neural network to satisfy this requirement.

\section{Machine-Learning Method}\label{sec:Method}

To accelerate the transformation to the target manifold $\tilde{\mathcal{M}}$, reducing computational complexity, it is possible to define a neural network trained to approximate the integrator~\eqref{eq:formal-integrator} $\NN\approx \Int^+_{\tau}$.

One approach is to learn the imaginary part of any configuration on the target manifold $\tilde{\mathcal{M}}$ given its real part\cite{PhysRevD.96.094505,leveragingML}
\begin{equation}
    \SHIFT: \M_0 \to \tilde{\M}, \, \Phi \mapsto \Phi + i NN\left( \Re{\Phi} \right).
    \label{eq:SHIFT}
\end{equation}
This ansatz has two advantages.
First, the ergodicity issue, induced by potential trapping on individual thimbles, is removed~\cite{PhysRevD.96.094505}.
Second, the network can use the well-established methods of real-valued neural networks.
Computational costs due to flowing are reduced as the application of the neural network is much cheaper then any numerical integration.
However, a major disadvantage is the computational effort and severe volume scaling of the Jacobian determinant~\cite{leveragingML}.

In this work we use complex-valued neural networks -- networks with complex parameters -- to instead learn the map from the integration manifold $\M_0$ to the target manifold $\tilde{M}$,
\begin{equation}
    \tilde{\Phi} = \NN(\phi) \approx \Int^+_\tau(\phi).
\end{equation}
This approach enjoys a significant advantage over the SHIFT network \eqref{eq:SHIFT}: given the right network architecture the Jacobian may be evaluated very quickly.
Below we will explain our use of affine coupling layers to reduce the scaling of the Jacobian determinant from a general cubic scaling down to a linear scaling in the volume $\abs{\Lambda}$. 

For a recent overview of complex-valued networks see \Ref{bassey2021survey}.
Typical automatic differentiation algorithms can be applied to complex-valued neural networks in a similar manner as to real-valued ones~\cite{bouboulis2010wirtinger,brandwood1983complex,kreutz2009complex} by switching the differentiation rule to Wirtinger derivatives~\cite{bouboulis2010wirtinger}
\begin{equation}
    \begin{aligned}
        \frac{\partial f(z)}{\partial z} &= \frac{1}{2} \left( \frac{\partial f(z)}{\partial \Re z} - i \frac{\partial f(z)}{\partial \Im z} \right)\\
        \frac{\partial f(z)}{\partial z^*} &= \frac{1}{2} \left( \frac{\partial f(z)}{\partial \Re z} + i \frac{\partial f(z)}{\partial \Im z} \right).
        \label{eq:Wirtinger-derivative}
    \end{aligned}
\end{equation}
The Wirtinger derivatives have the advantage that they coincide with complex derivatives for holomorphic functions  while also extending to non-holomorphic ones.
This generalization is required for two reasons.
First, loss functions typically are not holomorphic and are not differentiable in the complex sense.
Second, Liouville's theorem, stating that bounded entire functions are constant, reduces the usability of any complex-valued neural network if only holomorphic components can be used.
As automatic differentiation is possible through backpropagation using Wirtinger derivatives, these restrictions can be overcome and a neural network $\NN:\mathbb{C}^m \to \mathbb{C}^n$ with complex-valued weights can be defined \cite{bassey2021survey}.
We want to emphasize that a non holomorphic network can approximate the thimbles even though their definition is manifestly holomorphic. 
This can be understood by utilizing the universal approximation theorem~\cite{voigtlaender2020universal}, 
and realizing that the change of variable requires an embedding which is at least twice-differentiable in the Wirtinger sense.
It is expected that such networks have an improved expressivity compared to real valued networks of twice the size -- mimicking the real and imaginary parts -- as complex networks do not have to learn complex arithmetic~\cite{bassey2021survey}.

Special care has to be taken when evaluating the Jacobian induced by the parametrization of $\tilde{M}$.
The Jacobian in the effective action~\eqref{eq:effective-action} is defined by the derivative of the transformation according to its real parameters -- a derivative in the real sense.
When applying a non-holomorphic neural network to parametrize the manifold, the Wirtinger derivatives force us to reexpress the derivative in the real sense by combining the two equations of~\eqref{eq:Wirtinger-derivative} and the transformation on the tangent plane~\eqref{eq:tangent-plane}
\begin{align}
        J_{ij} &\equiv \frac{\partial \NN(\phi+ i \Phi^0_c)_i}{\partial \phi_j}
               &= \frac{\partial \NN(\Phi)_i}{\partial\Phi_j} + \frac{\partial \NN(\Phi)_i}{\partial\Phi_j^*}.
    \label{eq:Wirtinger-jacobian}
\end{align}

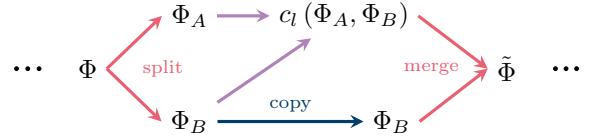
\begin{figure}
\begin{tikzpicture}
        \node[draw,circle, fill=black, inner sep = 0.5pt] at (0.0,0) {};
        \node[draw,circle, fill=black, inner sep = 0.5pt] at (0.15,0) {};
        \node[draw,circle, fill=black, inner sep = 0.5pt](dotsIn) at (0.3,0) {};

        \node[right of = dotsIn,xshift=-1em] (input) {$\Phi$};
        \node[right of = input,xshift=1em, yshift = 2em] (PhiA) {$\Phi_A$};
        \node[right of = input,xshift=1em, yshift =-2em] (PhiB) {$\Phi_B$};
        \node[right of = PhiA,xshift=3em] (c1) {$c_l\left(\Phi_A,\Phi_B\right)$};
        \node[right of = PhiB,xshift=4.8em] (PhiB copy) {$\Phi_B$};

        \node[right of = input, xshift =13em] (output) {$\tilde{\Phi}$};

        \draw[arrow,fzjred] (input.east) -- (PhiA.west);
        \draw[arrow,fzjred] (input.east) -- (PhiB.west);
        \node[right of = input, fzjred, font=\scriptsize] {split};

        \draw[arrow,fzjviolet] (PhiA) -- (c1);
        \draw[arrow,fzjviolet] (PhiB) -- (c1);
        \draw[arrow,fzjblue] (PhiB) -- node[above,font=\scriptsize] {copy} (PhiB copy);

        \draw[arrow, fzjred] (c1.east) --  (output.west);
        \draw[arrow, fzjred] (PhiB copy.east) -- (output.west);
        \node[left of = output, fzjred, font=\scriptsize] {merge};

        \node[draw,circle, fill=black, inner sep = 0.5pt,right of = output,xshift=-1em] {};
        \node[draw,circle, fill=black, inner sep = 0.5pt,right of = output,xshift=-0.6em] {};
        \node[draw,circle, fill=black, inner sep = 0.5pt,right of = output,xshift=-0.2em]{};

\end{tikzpicture}
    \caption{
        Pictorial representation of one coupling layer \eqref{eq:ACL-def}.
        First the input configuration $\Phi$ is split into two partitions $\Phi_A$ and $\Phi_B$.
        The corresponding $A$ components are then changed accoring to the prescribed coupling $c_l\left(\Phi_A,\Phi_B\right)$ 
        while the $B$ components are untouched.
        We utilize an affine transformation~\eqref{eq:affine-coupling} for the coupling $c_l$.
        The resulting output vector $\tilde{\Phi}$ is then constructed from the transformed $A$ components and unchanged $B$ components.
    } \label{fig:NN-architecture}
\end{figure}

To identify an architecture with an efficiently-computable Jacobian determinant, split the network into $L$ constituent layers:
\begin{align}
    \Phi_0(\phi)
    &=
    \phi + i \Phi_c^0
\nonumber
\\  \Phi_{\ell>0}(\phi)
    &=
    \NN_\ell(\Phi_{\ell-1}(\phi))
\nonumber
\\  &= \left(\NN_\ell \circ \NN_{\ell-1} \circ \cdots \circ \NN_1\right)(\phi)
\nonumber
\\  \tilde{\Phi}(\phi)  = \Phi_L(\phi)
    &= \NN_L(\phi)
    \equiv \NN(\phi).
\label{eq:layers}
\end{align}
The Jacobian determinant of the neural network\footnote{Note that this requires the input and output dimension of each layer to be equal.} is then given as the product of the Jacobian determinants of each layer
\begin{equation}
    \det{J} = \prod_{\ell=1}^{L} \det{J_{\NN_\ell}}.
\end{equation}
Consequently, we focus on layers with computationally simple Jacobian determinants.
Coupling layers
\begin{equation}
    \NN_\ell(\Phi) =
    \begin{cases}
        c_\ell\left[ \Phi_A, \, \Phi_B \right] & A_\ell \text{ components} \\
        \Phi_B & B_\ell \text{ components}
    \end{cases}
    \label{eq:ACL-def}
\end{equation}
fulfill this requirement~\cite{albergo2021introduction,foreman2021hmc}.
Here $A$ and $B$ are layer-specific partitions of the input vector $\Phi$ of equal cardinality $\half\abs{\Lambda}$, and $\Phi_{A,B}$ are the components of the input belonging to the indicated partition.
If the coupling layer $c_\ell\left[\Phi_A,\Phi_B\right]$ acts elementwise and is holomorphic in the components $\Phi_A$
\begin{equation}
   \frac{\partial c_\ell\left[\Phi_A,\Phi_B\right]}{\partial \Phi_A^*} = 0\,,
\end{equation}
the Jacobian determinant of each layer is given by
\begin{equation}
    \det{J_{\NN_\ell}(\Phi)} = \prod_{i=0}^{\abs{A}-1} \frac{\partial c_\ell\left[ \Phi_A, \Phi_B \right]}{\partial (\Phi_A)_i}\,.
\end{equation}
Furthermore, using an affine coupling~\cite{albergo2021introduction}
\begin{equation}
    c_\ell\left[\Phi_A, \Phi_B \right] = e^{m_\ell\left(\Phi_B\right)} \odot \Phi_A + a_\ell\left(\Phi_B\right)
    \label{eq:affine-coupling}
\end{equation}
with arbitrary differentiable functions $m_\ell,a_\ell: \mathbb{C}^{\nicefrac{\abs{\Lambda}}{2}} \to \mathbb{C}^{\nicefrac{\abs{\Lambda}}{2}}$ acting on the $B$ indices of the input configuration $\Phi$, yields a computationally cheap (log) Jacobian determinant
\begin{equation}
    \log{\det{J_{\NN}(\phi)}} = \sum_{\ell = 1}^{L} \sum_{i =0}^{\abs{A}-1} m_\ell\left( \Phi_{\ell-1}(\phi)_B\right)_i.
    \label{eq:logDetJ-NN}
\end{equation}
The expressivity of the neural network is controlled by the trainable parameters in the coupling functions $m_\ell, a_\ell$.
If $f$ denotes an affine transformation
\begin{equation}
    f(\Phi) = \omega \cdot \Phi + b
    \label{eq:affine}
\end{equation}
and $g$ the nonlinear ``softsign'' function
\begin{equation}
    g(z) = \frac{z}{1+\abs{z}}
    \label{eq:softsign}
\end{equation}
we take the coupling functions to be
\begin{equation}
    a_\ell,m_\ell =  g \circ f \circ g \circ f
    \label{eq:coupling functions}
\end{equation}
with independent complex weight matrices $\omega$ and bias vectors $b$.
The softsign function is non-holomorphic, requiring us to consider the Jacobian in the Wirtinger sense~\eqref{eq:Wirtinger-jacobian}.
Due to the structure of the Jacobian matrix, the nonzero non-holomorphic components $\partial_{\Phi_B^*} c_\ell$ do not contribute to the determinant \eqref{eq:logDetJ-NN}.
A graphical representation of this architecture is displayed in Figure~\ref{fig:NN-architecture}.

We add layers in pairs so that $L$ is even.
Each pair shares their partitioning.
In each pair the first layer modifies the $A$ partition \eqref{eq:ACL-def} and the next modifies the $B$ partition using the same ansatz with independent weights and biases.
Notice, the Jacobian determinant can be implemented so it is evaluated during the forward pass~\cite{albergo2021introduction} which reduces the required additional cost to only the sums of equation~\eqref{eq:logDetJ-NN}. 
Consequently, the Jacobian determinant in the effective action~\eqref{eq:effective-action} only adds a computational complexity linear in the volume $\abs{\Lambda}$.

The training setup was kept simple, allowing for further improvements in the future.
A standard $L_1$ loss function and the ADAM algorithm implemented in PyTorch~\cite{PyTorch} was used to train the network.
We kept the ADAM specific hyper parameters -- running average coefficients $\beta_i = (\num{0.9},\num{0.999})$, denominator shift $\varepsilon = \num{1e-08}$ as well as weight decay $w = 0$ -- at the standard values.
The training data comprised \num{10000} (\num{16000} for the 18 Sites) configurations drawn from normal distributions $\phi\sim \mathcal{N}_{0,\sigma}$, with $\sigma$ uniformly sampled between $\sqrt{\nicefrac{U}{\left(1+\nicefrac{16}{N_t}\right)}}$ and $\sqrt{U}$~\cite{leveragingML}, as input.
The ``labels'' consist of the corresponding flowed configurations $\Int^+_{\tau}(\phi)$, where the integration is performed using an adaptive Runge Kutta method of \nth{4} order.
A similar setup is used for the validation and testing data but only for $\num{2000}$ configurations each.
To avoid learning features of the thimbles irrelevant to the integral~\cite{leveragingML,kanazawa2015structure,PhysRevD.93.014504,mori2018lefschetz}, only configurations that did not flow to neverland are included in the training.

The network $\NN$ with 2 pairs of coupling layers was initialized to the identity so that before training it reproduced the tangent plane configurations which were fed into it.
We experimented with learning $\Int^+_\tau$ different flow times $\tau \in \{\num{1e-6},\,\num{1e-5},\,\num{1e-3},\,\num{1e-2},\,\num{1e-1}\}$.
We computed both the statistical power and measured correlators, as in Figure~\ref{fig:correlators}.
If we flow too much most configurations flow to neverland and training becomes expensive; if we flow too little the statistical power hardly improves.
The results shown in the next section have a flow-time $\tau = \num{1e-1}$.

Unfortunately picking a fixed flow time of this size was not feasable for the 18 sites problem. 
Instead, we defined a window of flow times $\tau \in [0,0.1]$ on which the flow is performed, as was originally done in~\cite{leveragingML}. 
In this manner, fixed flow-time configurations which would have flowed to neverland and thus have been rejected could still be used if their configurations remained valid within the window of flow times.
It was found in~\cite{leveragingML} that this method greatly decreased the cost generating training data.
In future work will continue to investigate more efficient ways of generating training data, and the training process itself, including by sampling one training point from the steps along a holomorphic flow to $\tau=\infty$ according to the real part of the step's action.

\section{Results}\label{sec:Results}

\begin{figure*}
    \includegraphics[width=\linewidth]{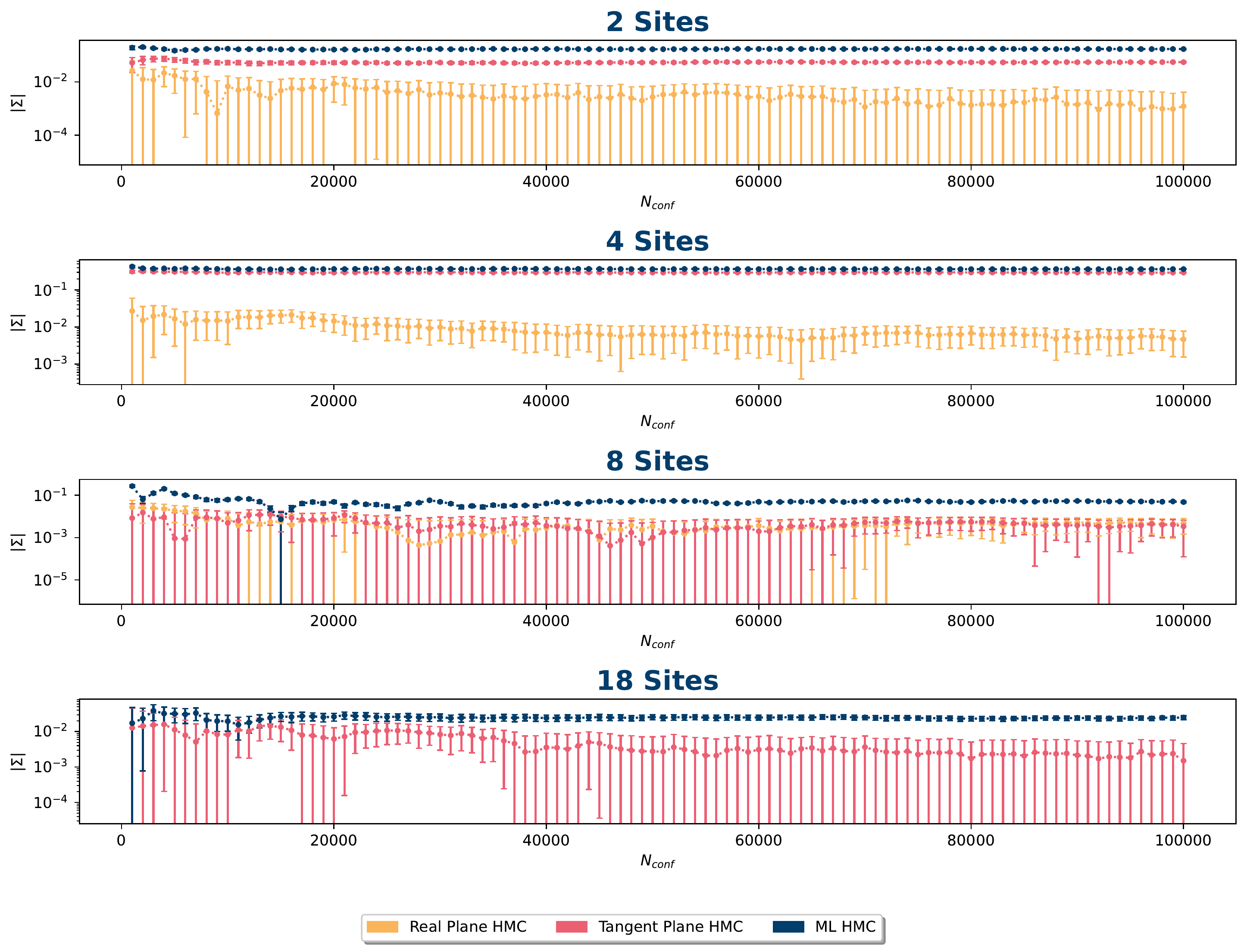}
    \caption{
        The statistical power $\abs{\Sigma} = \abs{\average{e^{-\mathrm{i} \Im{S}}}}$ is plotted against the number of configurations. 
        Three different algorithms are compared, the real plane (standard) HMC (orange), the tangent plane HMC (red) and the ML HMC (blue). 
        For 18 sites the real plane was totally noisy, and it is left out here.
        It can be seen that the ML HMC outperforms both real and tangent plane HMCs.
    }\label{fig:statpower}
\end{figure*}

We simulate the Hubbard model on the honeycomb lattices of 2, 4, 8 and 18 sites shown in \Figref{models}, using configurations obtained on the tangent plane and via our neural network \NN,
at inverse temperature $\beta=4$, $N_t = \num{32}$ timeslices, on-site coupling $U = \num{4}$, and chemical potential $\mu = \num{3}$.
To compare the machine learning enhanced HMC to other implementations such as the real-plane (standard) HMC with molecular dynamics on $\M_\Reals$ and the tangent plane HMC on $\M_T$ we consider the statistical power $\Sigma$.
A suitable algorithm will have $\abs{\Sigma}$ close to 1, whereas low values indicate a less suitable algorithm, since considerably more statistics would be required~\eqref{eq:effective-Nconf}.
Figure~\ref{fig:statpower} shows estimates of $\abs{\Sigma}$ with different numbers of configurations for the three mentioned HMC variants. The ML HMC is shown in blue, the tangent plane HMC in orange and the real plane HMC in red.
The ML HMC outperforms the two other algorithms in every case.
Moreover, in the case of 8 sites enormous statistics are required to even get a reasonable estimate of the statistical power for the real- and tangent-plane HMCs while the power of the ML HMC stabilizes with far fewer samples.
For 18 sites it was not feasible to simulate with the real plane HMC thus it is not shown here.
We can see that the tangent plane HMC does not get any reliable value for the statistical power while the MLHMC converges relatively fast.

\begin{figure*}[h]
    \includegraphics[width=\linewidth]{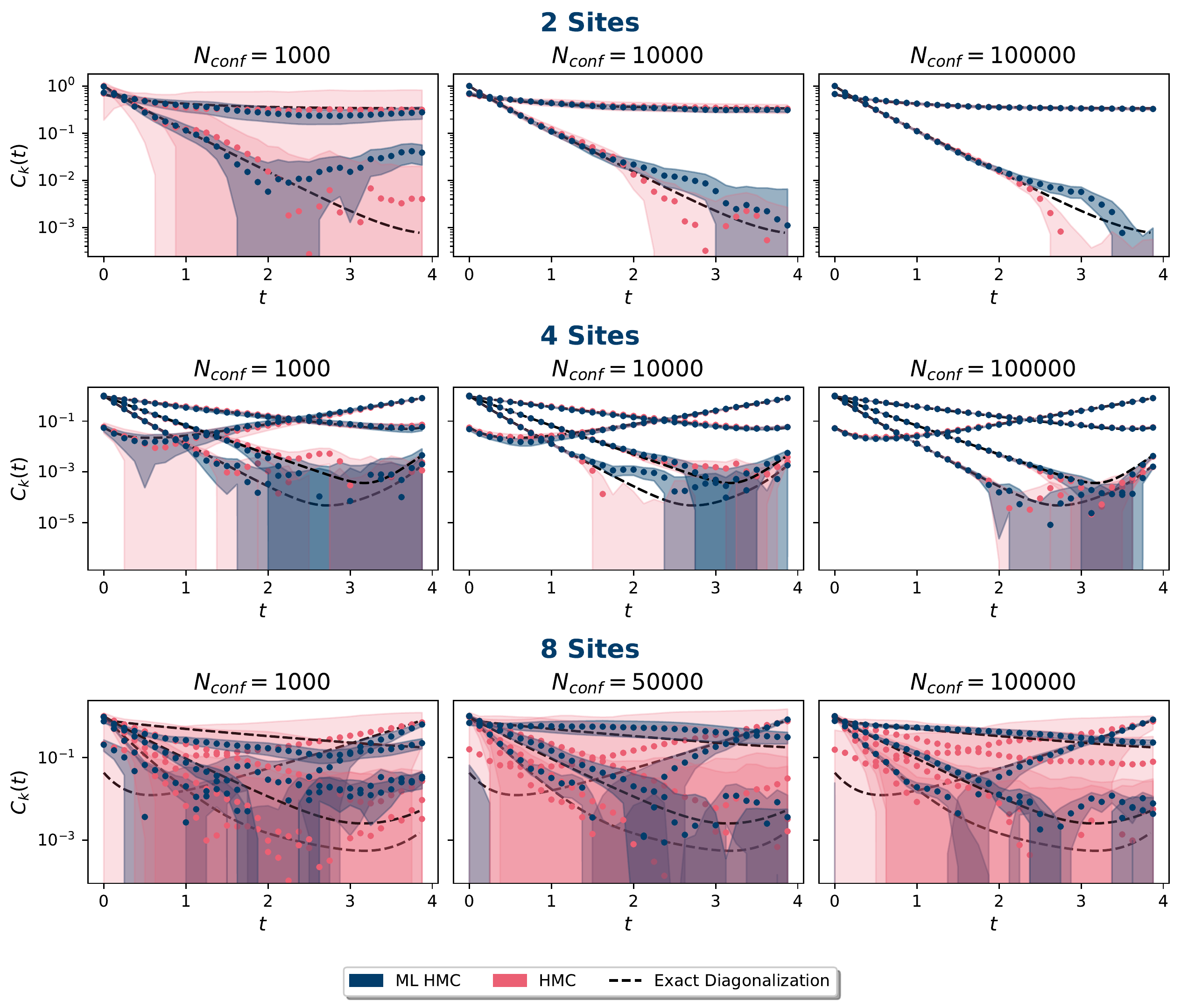}
    \caption{
        Momentum-projected correlation functions, measured using tangent plane HMC and ML HMC are shown in red and blue, respectively.
        These correlators were calculated with an inverse temperature $\beta=4$, $N_t = \num{32}$ time slices, on-site coupling $U = \num{4}$, and chemical potential $\mu = \num{3}$.
        The dashed black lines were determined by exact diagonalization.
        Each row corresponds to different number of ions (as in Fig.~\ref{fig:models}) increasing from top to bottom.
        Each column uses a different number of configurations $\Ncfg$ to estimate the correlators, increasing from left to right.
        Comparing the statistical power per $\Ncfg$ from figure~\ref{fig:statpower} suggests to use 
        $\Ncfg = \num{1000},\num{10000},\num{100000}$ for 2 and 4 sites as the uncertainty strongly differs.
        However, for 8 sites there is not much difference in the uncertainty of $\abs{\Sigma}$ between $\Ncfg = \num{1000}$ and $\num{10000}$;
        we show $\Ncfg = \num{1000}, \num{50000}, \num{100000}$ instead.
        ML HMC's improved statistical power shown in Fig.~\ref{fig:statpower} is reflected in the accuracy and uncertainty of these correlators.
        The sign problem of the 2 site and 4 site models is mild enough such that the tangent plane gives fairly good results, but the ML HMC gives more precise results with fewer configurations.
        For 8 sites the tangent plane HMC completely fails even at $\Ncfg = \num{100000}$ while ML HMC succeeds.
    }\label{fig:correlators}
\end{figure*}
\begin{figure*}
    \includegraphics[width = \linewidth]{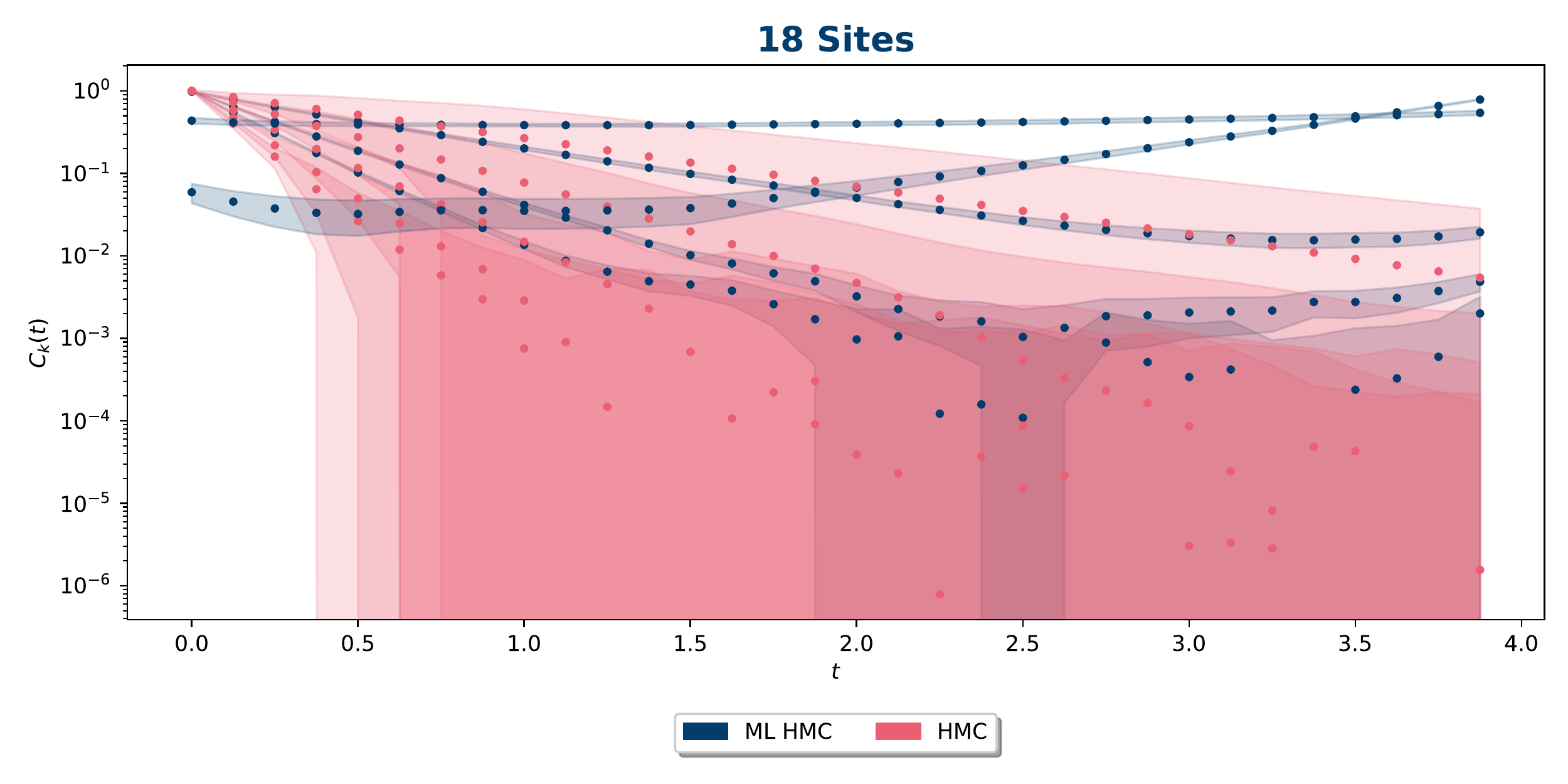}
    \caption{
        The single-particle correlators are displayed for the 18-site model at an
            inverse temperature $\beta = \num{4}$, with
            $N_t = \num{32}$ number of time slices, 
            with $U=\num{3}$ and $\mu=\num{3}$.
        The correlators have been measured with $\Ncfg = \num{100000}$ configurations.
        Again, the tangent plane HMC (red) does not provide any insight while the ML HMC resolves the correlators well.
        Assuming similarity to the smaller $U=4$ examples, ML HMC clearly determines the low-energy correlator while tangent plance HMC fails to find it at all.
    }\label{fig:18Sites-correlators}
\end{figure*}
\begin{figure*}
    \includegraphics[width=\linewidth]{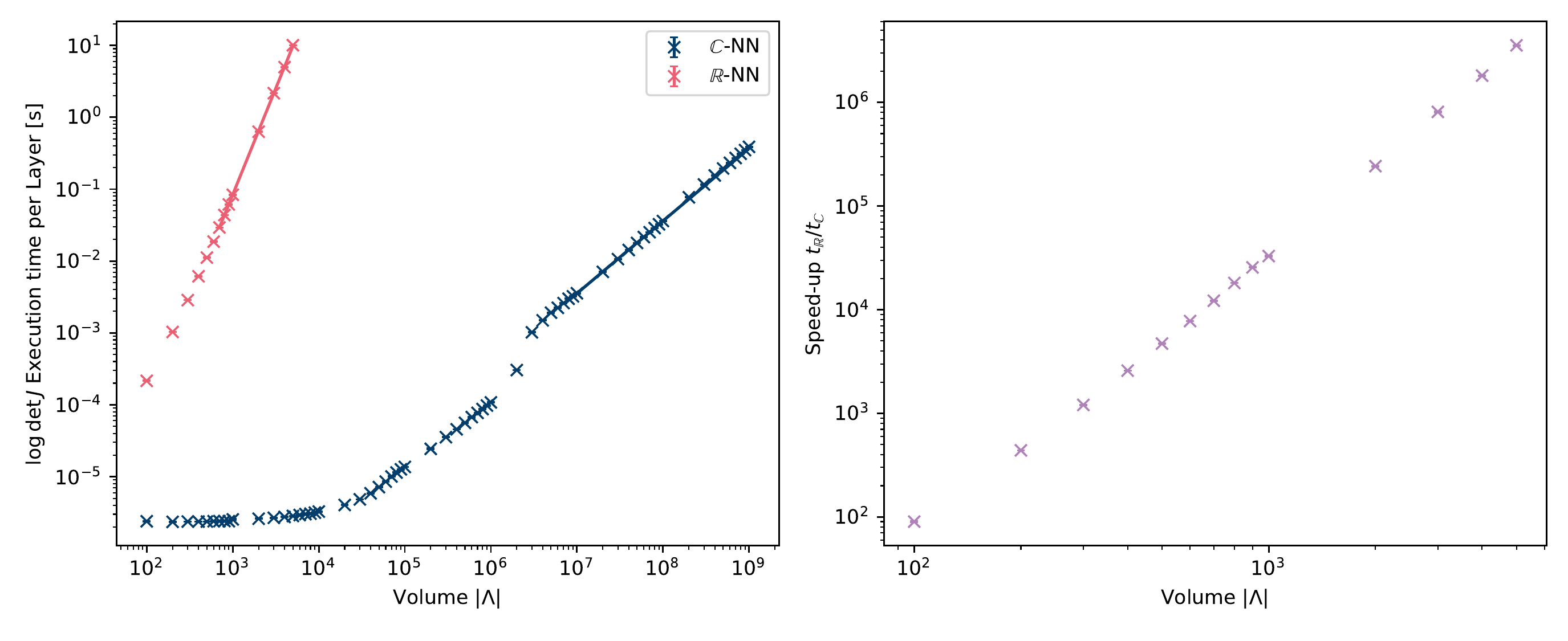}
    \caption{
        The left panel shows the scaling behavior per layer of $\log{\det{J}}$
        for the previously used SHIFT neural network, $\tilde{\Phi} = \Phi + i NN(\Phi)$ (red),
        and for the complex-valued paired affine coupling neural network, $\tilde{\Phi} = \NN(\Phi)$ (blue).
        On the right panel, we show the speedup for the different system volumes. 
        Theoretically the Jacobian determinant scaling of the SHIFT network is expected to be cubic in the system volume 
        while the $\NN$ is expected to scale linearly,
        resulting in a quadratic speed-up.
        The solid lines, on the left panel, represent log-log fits whose slopes determine the measured scaling orders.
        We find for the slopes of the SHIFT layer (red) a value of \num{2.955(1)} and for $\NN$ (blue) a value of \num{1.008(1)}, 
        resulting in a scaling improvement of power \num{1.947(2)}.
        The timing measurements were performed on JURECA~\cite{jureca} using one AMD EPYC CPU.
    }\label{fig:benchmark}
\end{figure*}

We show the efficacy of our method by computing euclidean-time correlators for a single particle or single hole created at time 0 and site $y$ and destroyed at time $t$ and site $x$.
\begin{align}
	C_{xy}^{p}(t) &= \average{ p_x^\dagger(t) p_y(0) } = \average{ M[+\Phi | +\mu ]\inverse_{xt;y0} }
\\
	C_{xy}^{h}(t) &= \average{ h_x^\dagger(t) h_y(0) } = \average{ M[-\Phi | -\mu ]\inverse_{xt;y0} }
\end{align}
To improve our signal we average on time slices in $t \in [\delta, \beta-\delta]$,
\begin{align}
	C_{xy}(t) &= \half \left( C_{xy}^p(t) + C_{xy}^{h*}(\beta-t) \right);
\end{align}
addends equal by symmetry even when $\mu\neq0$.
We then project both spatial indices to the same momentum $k$ to construct $C_k(t)$ for each momentum allowed by the lattice, and average correlators whose momenta are equal by rotational symmetry.

The match of our correlators in \Figref{correlators} with the exact results demonstrate that our algorithm is sampling the correct distribution.
Each row of the figure corresponds to one of the exactly-diagonalizable system sizes and each column restricts the number of configurations $\Ncfg$ used to estimate the correlators.
The red correlators are determined using a tangent plane HMC, the blue ones using ML HMC.
Finally, the black dashed lines correspond to the correlators obtained by an exact diagonalization procedure.
For the smaller examples the statistical errors of ML-HMC are much smaller, especially with fewer samples, as is expected from their respective statistical powers shown in \Figref{statpower}.
The worst sign problem can be found in the 8 sites case.
Here the tangent plane HMC fails even for $N_{conf} = \num{100000}$ and the statistical uncertainty in the correlators is essentially 100\%.
ML HMC obtains a weak signal at $N_{conf} = \num{50000}$ configurations and improves with greater statistics.

Finally, we compute correlators for a system with 18 sites and the same parameters but with $U = 3$ which is not tractable by exact diagonalization.
As shown in the statistical power plot \Figref{statpower} this model has a severe sign problem which could not be previously overcome.
Again comparing tangent plane and ML HMC in \Figref{18Sites-correlators} it can be seen that the ML HMC outperforms the tangent plane HMC and with the \num{100000} measurements quite a good signal is obtained.

In all cases we measured on every $10^{\textrm{th}}$ configuration such that no appreciable autocorrelation is found.
All these simulations indicate that the neural network improves the statistical power and uncertainty in observables quite drastically even when using a simple architecture.
We anticipate further improvements of our network by incorporating additional layers or incorporating knowledge of the problem's symmetries using equivariant layers~\cite{Favoni:2020reg,Luo:2020stn,Kanwar:2020xzo}.

The main advantage of our new complex architecture lies in the efficiency of the Jacobian determinant~\eqref{eq:logDetJ-NN} calculation.
The form of the determinant~\eqref{eq:logDetJ-NN} shows that it can be computed during the forward pass, reusing intermediate results from the application of the network, and is linear in the volume $\abs{\Lambda}^{\alpha_{\NN}}$
\begin{equation}
    \alpha_{\NN} = 1 \,.
    \label{eq:logDetJ-Order}
\end{equation}
The calculation of the determinant using a SHIFT layer~\cite{PhysRevD.96.094505,leveragingML} with the implementation of PyTorch~\cite{PyTorch}  (through LU-decomposition) scales with the third power of volume, $\abs{\Lambda}^{\alpha_{\SHIFT}}$ i.e. $\alpha_{\SHIFT} = 3$.
Measurements of the execution times of the determinant for the two neural network architectures are compared in Figure~\ref{fig:benchmark}.
The left panel shows the execution time per layer of $\log{\det{J}}$ for different artificial system volumes.
These volumes define the size of the configuration $\Phi$ which is randomly sampled and then passed to the networks.
On the log-log plot the linear behavior in the region $\abs{\Lambda}>\num{1e+7}$ -- for $\alpha_{\NN}$ -- and $\abs{\Lambda} > \num{7e+2}$ -- for $\alpha_\SHIFT$ -- determines the algorithms' scaling.
A simple least square fit provides the scaling exponents
\begin{equation}
    \begin{aligned}
        \alpha_{\SHIFT} &= \num{2.955(1)} \\
        \alpha_{\NN} &= \num{1.008(1)},
    \end{aligned}
\end{equation}
confirming our expected scaling behavior.
We then calculate the speedup achieved with the complex over the SHIFT network architecture in the right panel of figure~\ref{fig:benchmark}.
The expected quadratic speedup is confirmed by the benchmark result of
\begin{equation}
    \alpha_\SHIFT - \alpha_{\NN} = \num{1.947(2)}.
\end{equation}

\FloatBarrier
\section{Conclusions}
Mitigating the sign problem induced by a complex action is a major target of algorithmic development for simulating quantum-mechanical systems.
The application of neural networks approximating Lefschetz thimbles have shown great promise in the past.
We show that the supervised training of a simple complex-valued neural network architecture -- paired affine coupling layers with complex weights and biases -- allows for the successful simulation of systems with increasingly severe sign problems.
Our ML HMC approach reduces the sign problem sufficiently and enjoys a statistical power much greater than vanilla real-plane or tangent-plane HMC, 
as shown in Figure~\ref{fig:statpower},
improving the reliability of the correlator estimators in Figure~\ref{fig:correlators}.  We demonstrated the fidelity and correctness of our method by simulating 2, 4 and 8 site models and comparing our results to that obtained from direct diagonalization, obtaining excellent agreement.  We then applied our method to the 18 sites problem where direct diagonalization is not realizable.  Our results here thus represent predictions for this system in a regime where standard Monte Carlo methods are not possible due to the severity of the sign problem. 

Our results were made possible due to the favorable volume scaling of our new method.
Compared to previous methods we drastically reduced the computational cost of the Jacobian determinant from a general cubic scaling down to linear in the volume.
This has been numerically tested and demonstrated in Figure~\ref{fig:benchmark}.
Our computational complexity is therefore dominated by the application of the neural network itself,
and can be further improved by using sparse methods, convolutional layers, or other layer architectures. 
We are actively investigating such possibilities.
\begin{acknowledgments}
We thank Jan-Lukas Wynen for many helpful discussions.
This work was funded in part by the NSFC and the Deutsche Forschungsgemeinschaft (DFG, German Research
Foundation) through the funds provided to the Sino-German Collaborative
Research Center ``Symmetries and the Emergence of Structure in QCD''
(NSFC Grant No.~12070131001, DFG Project-ID 196253076 -- TRR110)
as well as the STFC Consolidated Grant ST/T000988/1.  
MR was supported under the RWTH Exploratory Research Space (ERS) grant PF-JARA-SDS005.
We gratefully acknowledge the computing time granted by the JARA Vergabegremium and provided on the JARA Partition part of the supercomputer JURECA at Forschungszentrum Jülich.
\end{acknowledgments}
 
\appendix

\bibliographystyle{apsrev4-1}

\end{document}